\documentclass[10pt,twocolumn,prl,aps,amssymb,amsmath,tightenlines,showpacs]{revtex4}
\usepackage{dsfont}
\usepackage{graphicx}
\usepackage{amssymb,amsfonts, overpic}

\newcommand{\half}{\mbox{$\textstyle \frac{1}{2}$}}

\newcommand{\re}{\mbox{$\rm e$}}
\newcommand{\ri}{\mbox{$\rm i$}}

\begin{document}

\title{Elementary solution to the time-independent quantum navigation problem}

\author{Dorje C. Brody$^{1,2}$ and David M. Meier$^1$}

\affiliation{$^1$Department of Mathematics, Brunel University, Uxbridge UB8 3PH, UK  \\ 
              $^2$St Petersburg State University of Information Technologies, Mechanics and Optics, \\ 
              Kronwerkskii ave 49, St Petersburg 197101, Russia }

\date{\today}

\begin{abstract}
A quantum navigation problem concerns the identification of a time-optimal Hamiltonian that 
realises a required quantum process or task, under the influence of a prevailing `background' 
Hamiltonian that cannot be manipulated. When the task is to transform one quantum state 
into another, finding the solution in closed form to the problem is nontrivial even in the case of time-independent 
Hamiltonians. An elementary solution, based on trigonometric analysis, is found here when 
the Hilbert space dimension is two. Difficulties arising from generalisations to higher-dimensional 
systems are discussed. 
\end{abstract}


\maketitle




Motivated in part by the advances in quantum technologies, significant progress has been made in 
finding the time-optimal scheme to implement a unitary operation that achieves the transformation of 
one quantum state into another, subject to a given set of constraints \cite{ML,lloyd1,brockett,brody1,lloyd3,
hosoya,brody2,hosoya2,zanardi,caneva,GCH,garon,stepney1,lloyd}. Typically it is assumed that 
external influences such as a background field or potential are absent, but in some cases it can be 
difficult to eliminate `ambient' Hamiltonians in a laboratory. In such a context, Russell 
\& Stepney \cite{stepney2} considered a time-minimisation problem of transporting one unitary 
operator ${\hat U}_I$ into another operator ${\hat U}_F$, subject to the existence of a background 
Hamiltonian ${\hat H}_0$ that cannot be manipulated. The task here therefore is to find the 
(time-dependent) control 
Hamiltonian ${\hat H}_1(t)$ such that ${\hat H}={\hat H}_0 + {\hat H}_1$ transforms ${\hat U}_I$ into 
${\hat U}_F$ in the shortest possible time. Evidently, there has to be a bound on the energy resource, 
which in their problem is given by the trace norm of the control---the 
`full throttle' condition: ${\rm tr}({\hat H}_1^2)=1$ at all time. In addition, to ensure the existence of viable controls it is assumed that 
the background Hamiltonian is not dominant, i.e. 
${\rm tr}({\hat H}_0^2)<1$. Inspired by the classical problem of navigation in the 
ocean in the presence of wind or currents \cite{zermelo,caratheodory}, this is referred to as the quantum 
Zermelo navigation problem \cite{stepney2}. The solution to this problem of 
constructing a unitary 
gate under an external field was obtained recently \cite{stepney3,brody3}, whereas the problem of 
finding the time-optimal transformation $|\psi_I\rangle \to |\psi_F\rangle$ of quantum states under a 
similar setup has not yet been solved. 

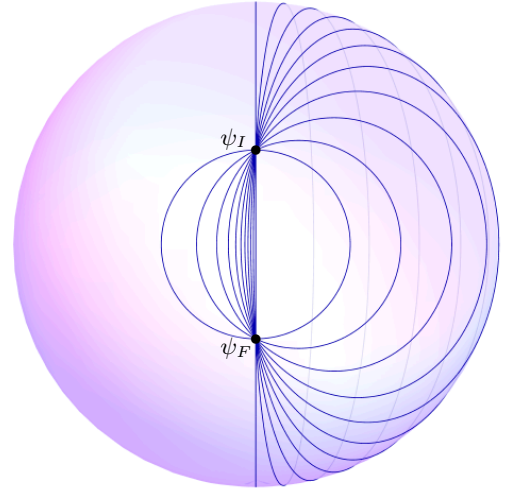
\begin{figure}[h!]
\begin{center}
\begin{overpic}[scale=0.40]{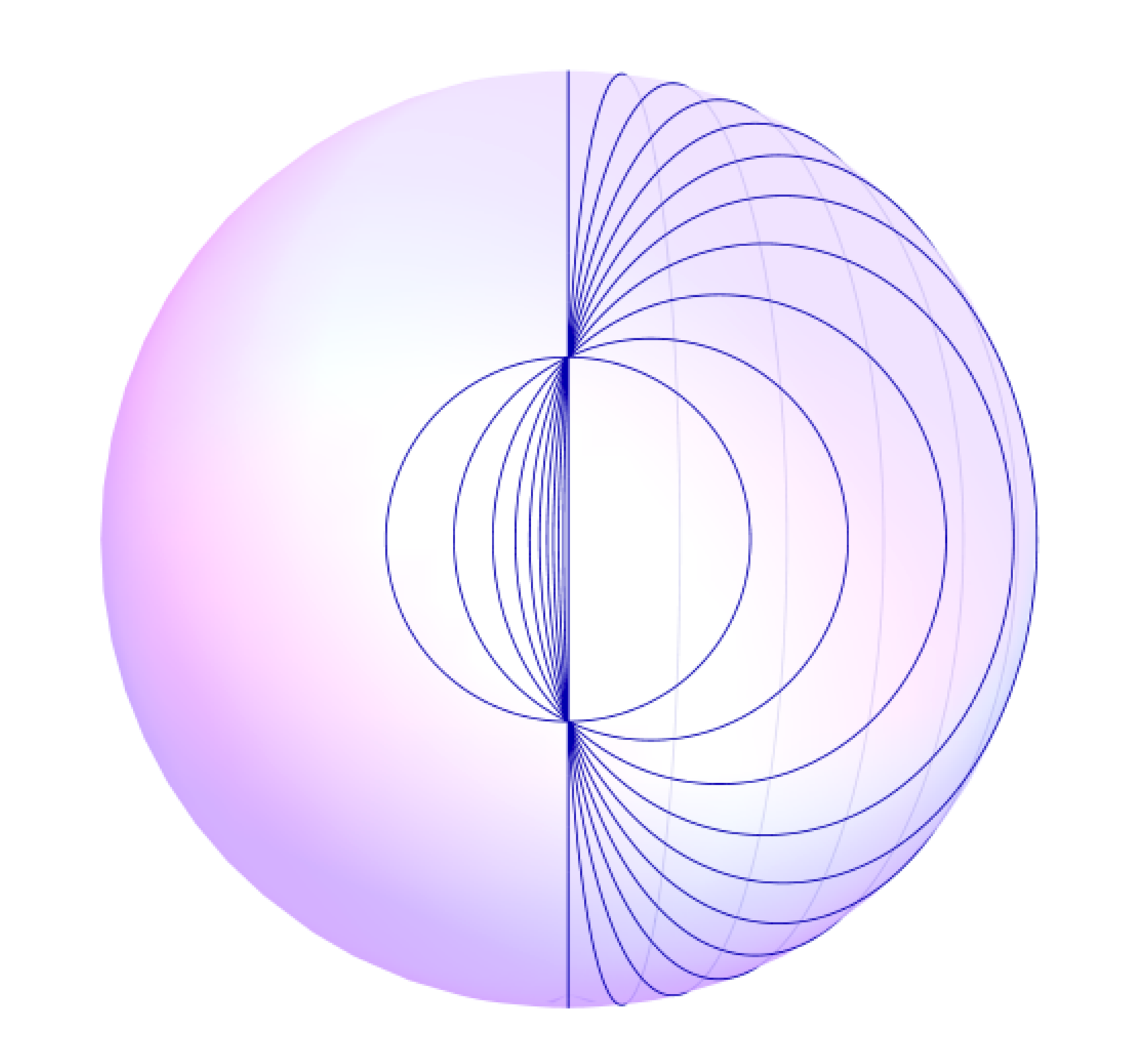}
\put(44.20, 63.5){$\psi_I$}
\put(49.2, 61.5) {$\bullet$}
\put(44.2, 27.60) {$\psi_F$}
\put(49.2, 29.2) {$\bullet$}
\end{overpic}
\end{center}
\footnotesize
\caption{\label{fig:0}
\textit{Circles on the sphere passing through a pair of points}. It is evident that the totality of latitudinal 
circles passing through the two points $|\psi_I\rangle$ and $|\psi_F\rangle$ share the property that 
the axes perpendicular to the circles lie on the plane that bisects all the circles. }
\end{figure}

In the present paper we investigate an analogous problem of finding the time-optimal control 
Hamiltonian ${\hat H}_1$ that achieves the transformation $|\psi_I\rangle \to |\psi_F\rangle$, 
subject to the existence of an ambient Hamiltonian ${\hat H}_0$, but in the time-independent 
context. It turns out that when ${\hat H}_1$ cannot vary in time, then the problem of finding the 
`time-optimal' control that generates a given unitary gate ${\hat U}_I \to {\hat U}_F$ 
becomes trivial (shown below), while that 
of finding the optimal control to generate the transformation $|\psi_I\rangle \to |\psi_F\rangle$ 
remains nontrivial. Nevertheless, in the case of two-level systems, on account of the fact that the 
configuration of the states can be `visualised' on a Bloch sphere, we are able to derive an 
elementary solution that requires nothing more than trigonometric manipulations. Our solution in fact 
extends to higher dimensions if the background Hamiltonian ${\hat H}_0$ (`wind') happens to leave 
invariant the Hilbert subspace spanned by the given two states $|\psi_I\rangle$ and 
$|\psi_F\rangle$; whereas the solution to the more general cases in higher dimensions remains 
open. 

\begin{figure}[t!]
\begin{center}
\vspace{-1.2cm}
\includegraphics[scale=0.38]{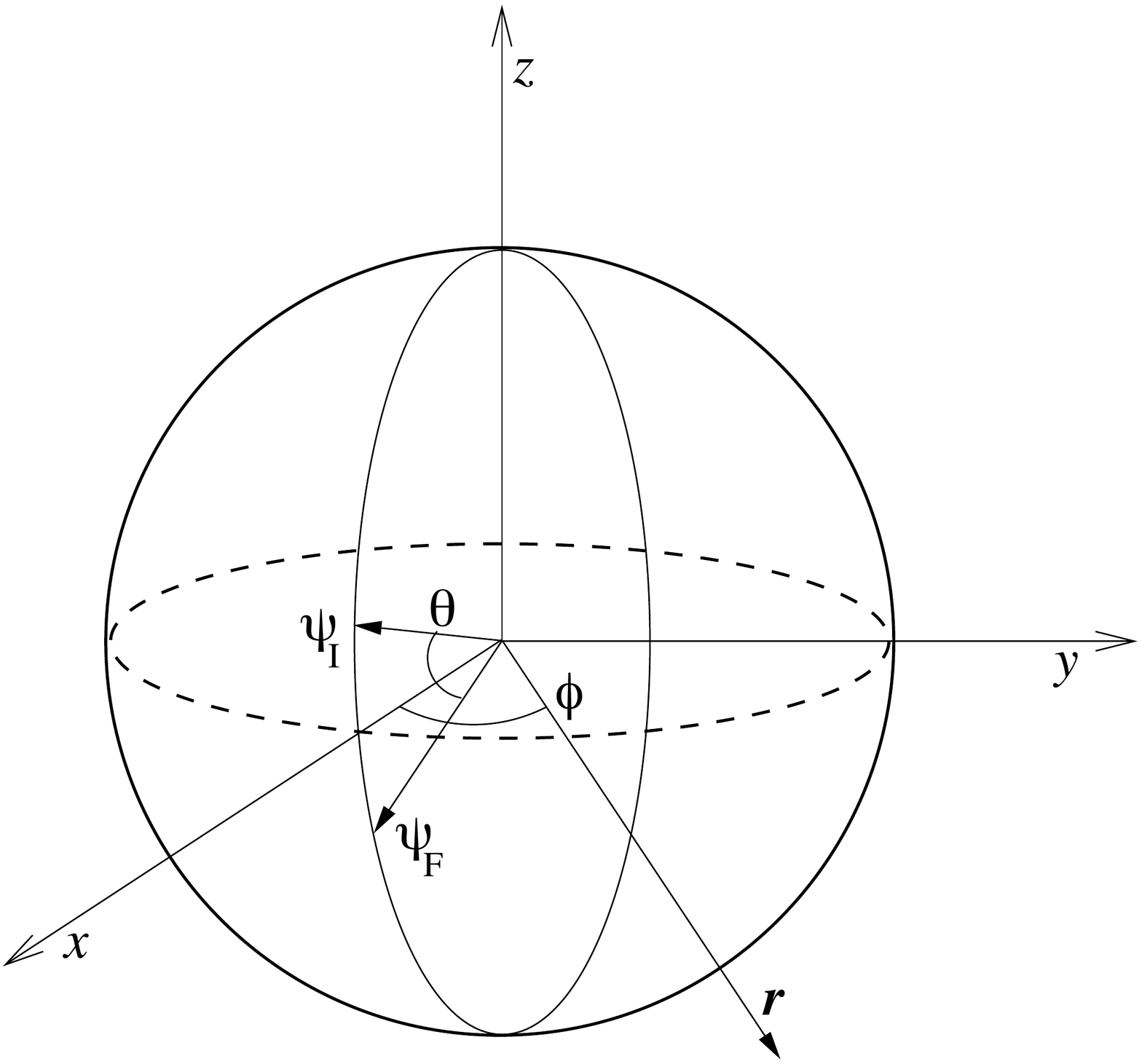}
\footnotesize
\vspace{-1.0cm}
\caption{\label{fig:1}
\textit{Configuration of the initial and target states}. Since the action of a time-independent 
Hamiltonian ${\hat H}={\boldsymbol r}\cdot{\hat{\boldsymbol \sigma}}$ amounts to a rigid rotation 
of the Bloch sphere, if the unitary operator $\re^{-{\rm i}{\hat H}t}$ were to transform a given 
initial state $|\psi_I\rangle$ into a target final state $|\psi_F\rangle$ at some time, the axis 
${\boldsymbol r}$ of rotation has to lie on the equator that bisects the great circle joining 
$|\psi_I\rangle$ and $|\psi_F\rangle$. When ${\boldsymbol r}$ points in the direction of the 
$y$-axis, the orbit $|\psi(t)\rangle=\re^{-{\rm i}{\hat H}t}|\psi_I\rangle$ is a geodesic curve, 
but owing to the prevailing `wind' ${\hat H}_0$, the journey along the shortest path does not result 
in the shortest time. }
\end{center}
\end{figure}

We begin our analysis by remarking that if a time-independent Hamiltonian ${\hat H}={\hat H}_0 + 
{\hat H}_1$ were to transform $|\psi_I\rangle$ into $|\psi_F\rangle$ in a two-dimensional Hilbert 
space, then since the action of ${\hat H}$ amounts to a rigid rotation of the associated Bloch sphere 
about some axis, the two states $|\psi_I\rangle$ and $|\psi_F\rangle$ must lie on the same latitudinal 
circle with respect to the axis of rotation determined by ${\hat H}$. A set of such circles is sketched in 
figure~\ref{fig:0}. Therefore, the totality of 
rotation axes permitting such transformations lie on the great circle passing the point 
$\frac{1}{\sqrt{2}}(|\psi_I\rangle+|\psi_F\rangle)$ that is orthogonal to the great circle joining the 
two points on the Bloch sphere corresponding to the states $|\psi_I\rangle$ and $|\psi_F\rangle$. 
Without loss of generality, let us work in the frame such that the two states can be expressed in the 
form 
\begin{eqnarray}
|\psi_I\rangle =  \left( \! \begin{array}{c} \cos \frac{1}{4}(\pi-\theta) \\ \\ 
\sin \frac{1}{4}(\pi-\theta) \end{array} \! \right) ,  \quad 
|\psi_F\rangle =  \left( \! \begin{array}{c} \cos \frac{1}{4}(\pi+\theta) \\ \\ 
\sin \frac{1}{4}(\pi+\theta) \end{array} \! \right) ,
\end{eqnarray}
where $\theta$ is the angular separation of the two states $|\psi_I\rangle$ and 
$|\psi_F\rangle$. In other words, we work with the coordinates such that both the initial 
and the target states lie on a longitudinal great circle, and such that the equator bisects the 
join of $|\psi_I\rangle$ and $|\psi_F\rangle$. By embedding the Bloch sphere in 
${\mathds R}^3$ we then find that the two points on the sphere corresponding to the 
vectors $|\psi_I\rangle$ and $|\psi_F\rangle$ lie on the $xz$-plane, located symmetrically about 
the $xy$-plane. Writing $\boldsymbol{\psi}_I$ and $\boldsymbol{\psi}_F$ for the two 
vectors in ${\mathds R}^3$ corresponding to the two states, we thus have 
\begin{eqnarray}
  \boldsymbol{\psi}_I = \frac{1}{2} 
  \left(\begin{array}{c} \cos \frac{\theta}{2} \\ 0\\ \sin \frac{\theta}{2} \end{array}\right), 
  \qquad \boldsymbol{\psi}_F = \frac{1}{2}
  \left(\begin{array}{c} \cos \frac{\theta}{2} \\ 0\\ -\sin \frac{\theta}{2} \end{array}\right), 
\end{eqnarray}
since the radius of the Bloch sphere is $\frac{1}{2}$. This configuration is schematically 
illustrated in figure~\ref{fig:1}. 

With the above choice of coordinates it should be evident that any rotation of the sphere 
about an axis that lies on the $xy$-plane will in time transport $|\psi_I\rangle$ into 
$|\psi_F\rangle$. Conversely, no rotation about an axis that does not lie on the $xy$-plane 
will ever transport $|\psi_I\rangle$ into $|\psi_F\rangle$. In the absence of the background 
`wind' ${\hat H}_0$, therefore, if the objective is to minimise the time subject to finite energy resource, 
then since the voyage time is the distance divided by speed, \textit{a priori} one has to deal 
with a complicated optimisation problem of minimising this ratio. However, 
fortunately in the case of a unitary evolution, the path that minimises the distance is 
precisely the path that maximises the evolution speed \cite{brody1}, so there is no need to evoke a 
simultaneous optimisation; all one needs is to find the shortest path. But geodesic curves on 
a sphere are given by the great circles, so without any calculation it is clear that the optimal 
Hamiltonian is given by the one corresponding to a rotation about the $y$-axis \cite{brody2}. 

In the presence of a background wind ${\hat H}_0$, however, the situation is different: 
In this case, depending on the choice of ${\hat H}$, that is, the choice of the rotation axis 
on the $xy$-plane, the energy resource available to the Hamiltonian ${\hat H}$ is different. 
As a consequence, one can find a Hamiltonian ${\hat H}$ such that although the path 
$|\psi(t)\rangle$ joining $|\psi_I\rangle$ and $|\psi_F\rangle$ is not the shortest, there is 
sufficient energy resource to overcome the extra mileage such that the voyage time will be 
shorter than that corresponding to the rotation about the $y$-axis. The objective, therefore, 
is to find the axis for which the voyage time is minimised. 

With these observations at hand, let us write the background Hamiltonian ${\hat H}_0$ in 
the form 
\begin{eqnarray}
{\hat H}_0 = \sqrt{\frac{\epsilon}{2}} \big( x {\hat\sigma}_x+y {\hat\sigma}_y+z 
{\hat\sigma}_z \big) , 
\end{eqnarray} 
where $x^2+y^2+z^2=1$ and where $0<\epsilon<1$. It follows that ${\rm tr}({\hat H}_0^2) 
= \epsilon<1$. Whatever the control Hamiltonian ${\hat H}_1$ might be, the total Hamiltonian 
has to take the form  
\begin{eqnarray}
{\hat H} = \frac{\omega}{2} \big( \cos\phi \, {\hat\sigma}_x+ \sin\phi \, {\hat\sigma}_y \big) 
\label{eq:4} 
\end{eqnarray} 
for some $\omega$ satisfying the constraint. 
In other words, the axis of rotation generated by ${\hat H}$ is at some angle $\phi$ from the 
$x$-axis on the $xy$-plane. Since ${\hat H}_1={\hat H}-{\hat H}_0$, the constraint 
${\rm tr}({\hat H}_1^2)=1$ on the control Hamiltonian implies that  
\begin{eqnarray}
\omega^2 - 2 \sqrt{2\epsilon} (x\cos\phi+y\sin\phi)\omega -2(1-\epsilon)=0 . 
\label{eq:5} 
\end{eqnarray}
We shall find that the voyage time $\tau$ such that the condition 
$\re^{-{\rm i}{\hat H}\tau} |{\psi}_I\rangle = |{\psi}_F\rangle$ is met 
will also depend on the variables $\omega$ and $\phi$. 
Thus, our objective is to minimise $\tau$ subject to the constraint (\ref{eq:5}). 

It should be remarked parenthetically that we have chosen both ${\hat H}_0$ and 
${\hat H}_1$ be trace free. This is because a physically meaningful constraint on the 
energy resource, in the case of a quantum system modelled on a finite-dimensional 
Hilbert space, is linked to the gap between the highest and the lowest attainable energy 
eigenvalues, not to the value of the ground-state energy \cite{brody1}. We shall therefore 
be working, without loss of generality, with trace-free Hamiltonians. 

\begin{figure}[t]
\begin{center}
\includegraphics[scale=0.20]{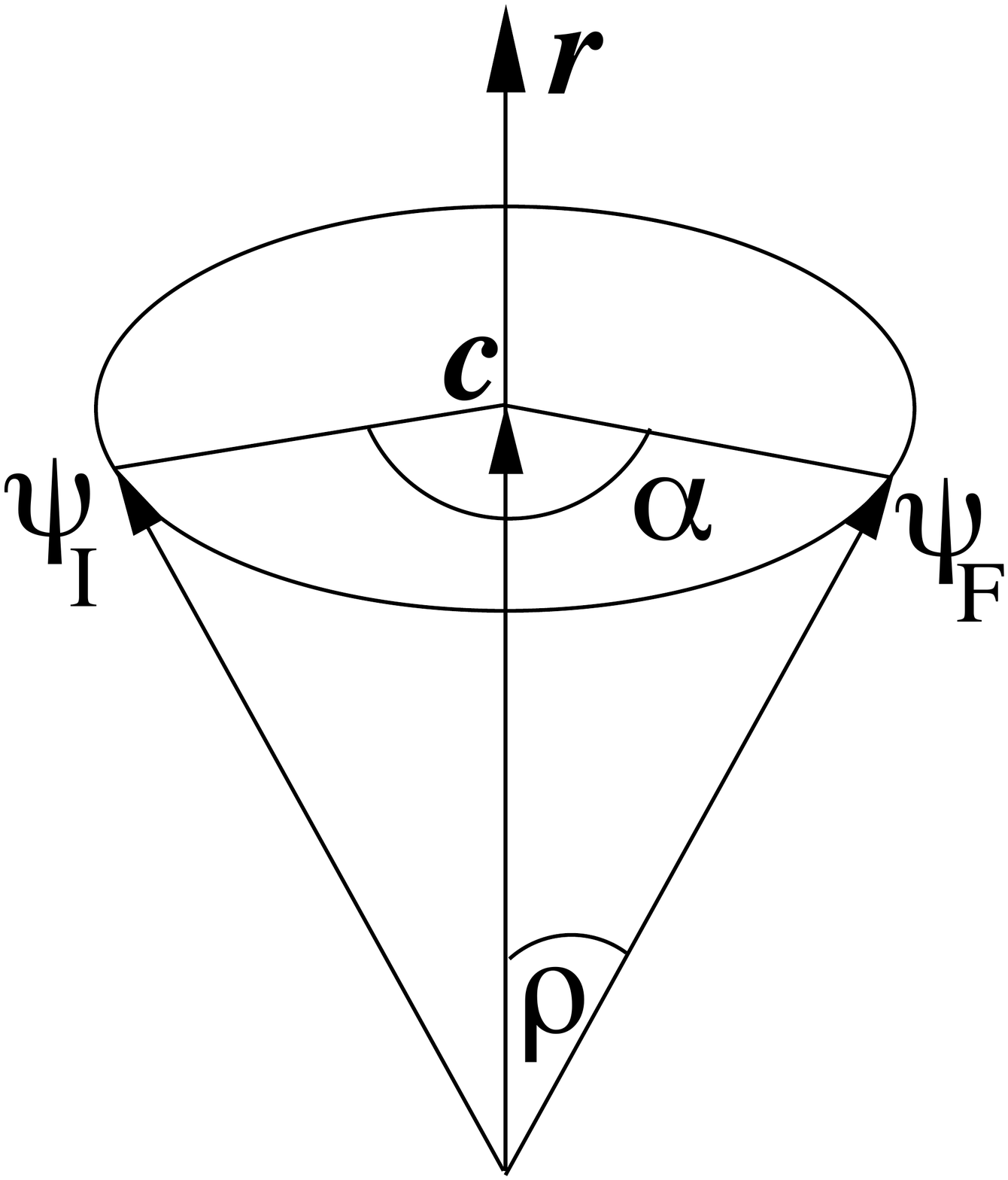}
\footnotesize
\caption{
\label{fig:2}
\textit{Identification of the rotation angle}. The angle $\alpha$ of rotation about the 
${\boldsymbol r}$-axis required to turn the vector ${\boldsymbol\psi}_I$ into 
${\boldsymbol\psi}_F$ is determined by first identifying the vector ${\boldsymbol c}$ 
such that ${\boldsymbol\psi}_I-{\boldsymbol c}\perp{\boldsymbol r}$. Then we have 
$({\boldsymbol\psi}_I-{\boldsymbol c})\cdot ({\boldsymbol\psi}_F-{\boldsymbol c}) 
\propto \cos\alpha$, from which $\alpha$ can be obtained.}
\end{center}
\end{figure}

To proceed, it should be evident from the foregoing formulation that the voyage time is 
proportional to the angle, call it $\alpha$, of rotation about the ${\hat H}$-axis that turns 
the vector $\boldsymbol{\psi}_I$ into $\boldsymbol{\psi}_F$ in ${\mathds R}^3$. Specifically, 
since the angular frequency generated by the Hamiltonian ${\hat H}$ of (\ref{eq:4}) is 
$\omega$, this in turn determines the voyage time according to the relation $\alpha=
\omega t$. It follows that 
the problem reduces to working out elementary trigonometric relations. Let us define the 
vector ${\boldsymbol r}$ by 
\begin{eqnarray}
{\boldsymbol r} =  \frac{\omega}{2} 
\left(\begin{array}{c} \cos \phi \\ \sin \phi \\ 0 \end{array}\right)
\label{eq:6} 
\end{eqnarray}
so that ${\hat H}={\boldsymbol r}\cdot{\hat{\boldsymbol \sigma}}$. Thus ${\boldsymbol r}$ 
determines the axis of rotation in ${\mathds R}^3$ generated by ${\hat H}$. To determine 
$\alpha$, let us first identify the angular separation $\rho$ between ${\boldsymbol r}$ and 
$\boldsymbol{\psi}_I$ (which, of course, is the same as that between ${\boldsymbol r}$ 
and $\boldsymbol{\psi}_F$ on account of the symmetry). To assist the analysis, in 
figure~\ref{fig:2} we give the 
perspective of the configuration around the ${\boldsymbol r}$-axis. Since 
${\boldsymbol r}\cdot\boldsymbol{\psi}_I=|{\boldsymbol r}|\, |\boldsymbol{\psi}_I|\, \cos\rho$, 
we find 
\begin{eqnarray}
 \cos \rho =  \cos \phi \cos \half \theta . 
\label{eq:7}
\end{eqnarray}
The final step required is to identify the vector ${\boldsymbol c}$ depicted in figure~\ref{fig:2} 
that points in the direction of ${\boldsymbol r}$ such that the two points 
$\boldsymbol{\psi}_I$ and $\boldsymbol{\psi}_F$ lie on the plane perpendicular to 
${\boldsymbol r}$ at ${\boldsymbol c}$. But clearly this is given by 
\begin{eqnarray}
{\boldsymbol c} = \half \cos\rho 
\left(\begin{array}{c} \cos \phi \\ \sin \phi \\ 0 \end{array}\right) , 
\end{eqnarray}
from which it follows, after some algebra, that 
\begin{eqnarray} 
\cos \alpha &=& \frac{\boldsymbol{\psi}_I - {\boldsymbol c}}{|\boldsymbol{\psi}_I - {\boldsymbol c}|} 
\cdot \frac{\boldsymbol{\psi}_F - {\boldsymbol c}}{
|\boldsymbol{\psi}_F - {\boldsymbol c}|} \nonumber \\ &=& 
\frac{\sin^2 \phi - \tan^2 \frac{\theta}{2}}{\sin^2 \phi + \tan^2 \frac{\theta}{2}} . 
\end{eqnarray}
As indicated above, since the angular frequency is 
$\omega$, the first time $\tau$ at which the state $|\psi_I\rangle$ is turned into $|\psi_F\rangle$ 
is given by $\omega\tau=\alpha$, that is, 
\begin{eqnarray} 
\tau = \frac{1}{\omega}\, \cos^{-1}\left( 
\frac{\sin^2 \phi - \tan^2 \frac{\theta}{2}}{\sin^2 \phi + \tan^2 \frac{\theta}{2}}\right) .
\label{eq:10}
\end{eqnarray} 
On the other hand, the constraint (\ref{eq:5}) allows us to express $\omega$ in terms of 
$\phi$. Putting these together, the first voyage time $\tau=\tau(\phi)$ can be expressed 
explicitly as a function of the angle $\phi$ that determines the axis of rotation generated 
by ${\hat H}$, which in turn determines ${\hat H}_1$. Specifically, we have 
\begin{eqnarray} 
\omega &=& \sqrt{2\epsilon(x\cos\phi+y\sin\phi)^2+2(1-\epsilon)}\nonumber \\ && \quad 
+\sqrt{2\epsilon}(x\cos\phi+y\sin\phi) , 
\label{eq:11}
\end{eqnarray} 
which together with (\ref{eq:10}) gives $\tau(\phi)$, and this in turn must be minimised 
for fixed $x$, $y$, $\epsilon$ and $\theta$. 

\begin{figure}[t]
\begin{center}
\includegraphics[scale=0.65]{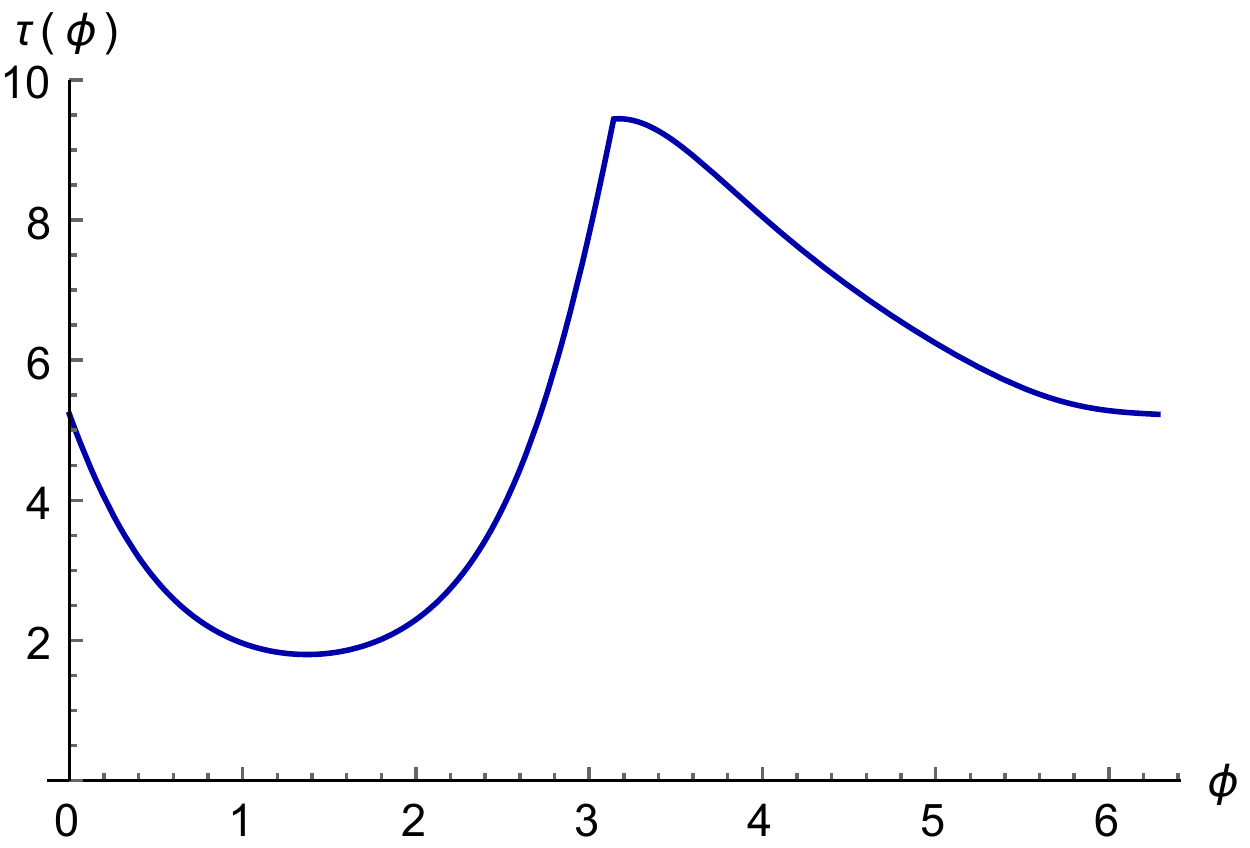}
\footnotesize
\caption{
\label{fig:3}
\textit{Voyage time $\tau(\phi)$ as a function of the angle $\phi$}. As the axis of rotation 
generated by ${\hat H}$, parameterised by $\phi$, is varied, the voyage time changes 
accordingly. Here, as an example, we plot $\tau(\phi)$ for the parameter choice: 
$x=0.1$, $y=0.23$, $z=(1-x^2-y^2)^{\frac{1}{2}}\approx0.97$, $\epsilon=0.9$, and 
$\theta=\pi/2$. In this example, the optimal angle is $\phi^*\approx 0.44\, \pi$. At $\phi=\pi$ 
there is a cusp (irrespective of the parameter values), since the orientation of the path
$|\psi(t)\rangle=\re^{-{\rm i}{\hat H}t}|\psi_I\rangle$ changes as $\phi$ passes through 
$\pi$. }
\end{center}
\end{figure}

In figure~\ref{fig:3} we plot $\tau(\phi)$ as a function of $\phi$ for a choice of parameters 
$x$, $y$, $\epsilon$ and $\theta$. Since the problem is reduced to a one-dimensional 
minimisation task, the optimal value $\phi^*$ for the axis of rotation can easily be determined 
numerically, which, when substituted in (\ref{eq:11}) and in (\ref{eq:4}), identifies the 
optimal overall Hamiltonian ${\hat H}(\phi^*)$, from which the optimal control can be 
determined by the relation ${\hat H}_1(\phi^*)={\hat H}(\phi^*)-{\hat H}_0$. This completes our 
analysis of finding the time-optimal Hamiltonian that generates the transformation 
$|\psi_I\rangle \to |\psi_F\rangle$ of quantum states, subject to the existence of a prevailing 
`wind' ${\hat H}_0$. As for the time required to achieve the transformation, this is given by 
$\tau(\phi^*)$. 

We conclude by remarking on the generalisation to higher dimensions, as well as on the 
problem of optimally generating a given unitary gate. In either case, in the time-independent 
context the evolution operator is given by $\re^{-{\rm i}{\hat H}T}$ for some Hamiltonian 
${\hat H}$ and voyage time $T$. Thus the task is to find the best choice of 
${\hat H}$ that minimises $T$ such that either 
\begin{eqnarray}
\re^{-{\rm i}{\hat H}T} |{\psi}_I\rangle = |{\psi}_F\rangle 
\label{eq:12}
\end{eqnarray} 
or 
\begin{eqnarray}
\re^{-{\rm i}{\hat H}T} {\hat U}_I = {\hat U}_F , 
\label{eq:13}
\end{eqnarray} 
is realised, depending on which problem one is considering. Now for a system modelled 
on an $n$-dimensional Hilbert space, the space of pure states is the associated projective 
Hilbert space of $n-1$ complex dimensions. Thus, the specification of a state requires the 
specification of $2n-2$ degrees of freedom. On the other hand, the specification of a 
Hamiltonian, up to trace, requires $n^2-1$ degrees of freedom. Together with the fact that 
$T$ is also unknown, we have, in (\ref{eq:12}), $n^2$ unknowns; while, noting 
that there is also the trace-norm condition, there are $(2n-2)+1=2n-1$ conditions. It follows 
that the solution to the problem of the type (\ref{eq:12}) involves an optimisation over 
$n^2-(2n-1)=(n-1)^2$ parameters, which in general is nontrivial. For $n=2$, this reduces 
to a single-parameter optimisation, and 
an explicit representation of $\tau$ in terms of trigonometric functions can be found, as shown 
above. For $n>2$, our solution remains valid if ${\hat H}_0$ leaves invariant the 
two-dimensional Hilbert subspace spanned by $|{\psi}_I\rangle$ and $|{\psi}_F\rangle$, 
on account of the observation made in \cite{brody2}; whereas in the general case, the 
voyage time $\tau$ will depend on $(n-1)^2$ parameters, hence a numerical search in a 
higher-dimensional parameter space is required to identify the optimal $\tau^*$. It remains 
open whether a similarly simple analytical form of $\tau$ can be found in higher dimensions. 
In any event, the problem of the kind represented in (\ref{eq:12}) is in general nontrivial, even 
in the time-independent context. As for the construction of a unitary gate as in (\ref{eq:13}), on 
the other hand, the situation is markedly different. Here the number of unknowns remains the 
same, but the number of constraints in (\ref{eq:13}), together with the trace condition, 
completely counterbalances this (recall that while (\ref{eq:12}) is a vector relation, (\ref{eq:13}) 
is a matrix relation), and there is no degree of freedom left to optimise. That is, the optimal 
Hamiltonian is given exactly by ${\hat H}^* = \ri T^{-1} \ln({\hat U}_F {\hat U}_I^{-1})$, where 
$T$ is fixed by the trace-norm condition. Specifically, writing ${\hat X}=\ri \ln({\hat U}_F 
{\hat U}_I^{-1})$ for simplicity, we have, on account of ${\rm tr}({\hat H}_1^2)=
{\rm tr}((T^{-1}{\hat X}-{\hat H}_0)^2)=1$, 
\begin{eqnarray}
\frac{1}{T} = \frac{ \sqrt{({\rm tr}({\hat H}_0{\hat X}))^2+[1-{\rm tr}({\hat H}_0^2)] 
{\rm tr}({\hat X}^2)} + {\rm tr}({\hat H}_0{\hat X}) 
} {{\rm tr}({\hat X}^2)}  \, \, \, 
\label{eq:14} 
\end{eqnarray}
for the voyage time required to realise the transformation ${\hat U}_I \to {\hat U}_F$. Thus, in 
the case of time-independent Hamiltonians, the problem of finding a time-optimal Hamiltonian 
to generate a unitary gate, under the influence of a background Hamiltonian ${\hat H}_0$, is 
empty---only with a time-dependent control ${\hat H}_1(t)$ the `bound' in (\ref{eq:14}) can be 
overcome \cite{stepney3,brody3}.




\begin{thebibliography}{}

\bibitem{ML} 
Margolus,~N. \& Levitin,~L.~B. 1998 
The maximum speed of dynamical evolution. 
{\em Physica} D\textbf{120}, 188.

\bibitem{lloyd1} 
Lloyd,~S. 1999 
Ultimate physical limits to computation. 
{\em Nature} \textbf{406}, 1047.

\bibitem{brockett} 
Khaneja,~N., Glaser,~S.~J. \& Brockett,~R. 2002 
Sub-Riemannian geometry and time optimal control of three spin systems: 
Quantum gates and coherence transfer. 
{\em Phys. Rev.} A\textbf{65}, 032301. 

\bibitem{brody1} 
Brody,~D.~C. 2003 
Elementary derivation for passage time. 
{\em J. Phys.} A\textbf{36}, 5587.  

\bibitem{lloyd3} 
Giovannetti,~V., Lloyd,~S. \& Maccone,~L. 2003 
Quantum limits to dynamical evolution. 
{\em Phys. Rev.} A\textbf{67}, 052109. 

\bibitem{hosoya} 
Carlini,~A., Hosoya,~A., Koike,~T. \& Okudaira,~Y. 2006 
Time-optimal quantum evolution. 
{\em Phys. Rev. Lett.} \textbf{96}, 060503. 

\bibitem{brody2} 
Brody,~D.~C. \& Hook,~D.~W. 2006 
On optimum Hamiltonians for state transformations. 
{\em J. Phys.} A\textbf{39}, L167. 

\bibitem{hosoya2} 
Carlini,~A., Hosoya,~A., Koike,~T. \& Okudaira,~Y. 2007 
Time-optimal unitary operations. 
{\em Phys. Rev.} A\textbf{75}, 042308.

\bibitem{zanardi} 
Rezakhani,~A.~T., Kuo,~W.~J., Hamma,~A., Lidar,~D.~A. \& Zanardi,~P. 2009 
Quantum adiabatic brachistochrone. 
{\em Phys. Rev. Lett.} \textbf{103}, 080502. 

\bibitem{caneva} 
Caneva,~T., Murphy,~M., Calarco,~T., Fazio,~R., Montangero,~S., Giovannetti,~V. 
\& Santoro,~G.~E. 2009 
Optimal control at the quantum speed limit. 
{\em Phys. Rev. Lett.} \textbf{103}, 240501. 

\bibitem{GCH} 
Hegerfeldt,~G.~C. 2013 
Driving at the quantum speed limit: Optimal control of a two-level system. 
{\em Phys. Rev. Lett.} \textbf{111}, 260501. 

\bibitem{garon} 
Garon,~A., Glaser,~S.~J. \& Sugny,~D. 2013 
Time-optimal control of SU(2) quantum operations. 
{\em Phys. Rev.} A\textbf{88}, 043422. 

\bibitem{stepney1} 
Russell,~B. \& Stepney,~S. 2013 
Geometric methods for analysing quantum speed limits: Time-dependent controlled quantum 
systems with constrained control functions. 
In {\em Unconventional Computation and Natural Computation}, G.~Mauri, \textit{et al}. Eds., 
(Berlin: Springer-Verlag). 

\bibitem{lloyd} 
Wang,~X., Allegra,~A., Jacobs,~K., Lloyd,~S., Lupo,~C. \& Mohseni,~M. 2014 
Quantum brachistochrone curves as geodesics: obtaining accurate control protocols 
for time-optimal quantum gates. arXiv:1408.2465 

\bibitem{stepney2} 
Russell,~B. \& Stepney,~S. 2014 
Zermelo navigation and a speed limit to quantum information processing. 
{\em Phys. Rev.} A\textbf{90}, 012303. 

\bibitem{zermelo} 
Zermelo,~E. 1931
\"Uber das Navigationsproblem bei ruhender oder ver\"anderlicher Windverteilung. 
{\em Ztschr. f. angew. Math. und Mech.} \textbf{11}, 114. 

\bibitem{caratheodory} 
Carath\'eodory,~C. 1935 
{\em Variationsrechnung und Partielle Differentialgleichungen erster Ordnung}. 
(Berlin: B.~G.~Teubner). 

\bibitem{stepney3} 
Russell,~B. \& Stepney,~S. 2014  
Zermelo navigation in the quantum brachistochrone. 
arXiv:1409.2055  

\bibitem{brody3} 
Brody,~D.~C. \& Meier,~D.~M. 2014 
Solution to the quantum Zermelo navigation problem. 
arXiv:1409.3204







\end{thebibliography}
\end{document}